\date{}
\documentclass[12pt,a4paper]{article}
\usepackage[affil-it]{authblk}
\usepackage{cancel}
\usepackage[utf8x]{inputenc}
\usepackage{multirow}
\usepackage{xfrac}
\usepackage{floatrow}
\usepackage[font=footnotesize, labelfont={sf,bf},margin=0cm]{caption}
\usepackage{lipsum}
\usepackage{textcomp}
\usepackage[english]{babel}
\usepackage[margin=1in, paperwidth=8.5in, paperheight=11in]{geometry}
\usepackage{amsmath}
\usepackage{graphicx}
\usepackage{epstopdf}
\usepackage{setspace}
\newfloat{suppfigure}{tbh}{losf}
\floatname{suppfigure}{Figure}
\makeatletter 
\renewcommand{\thesuppfigure}{S\@arabic\c@suppfigure}
\makeatother
\begin{document}
\title{Mathematical Modeling of CRISPR-CAS system effects on biofilm formation}
\author{Qasim Ali\thanks{Email address: \texttt{qali6@uwo.ca}} , Lindi Wahl}
\affil{Department of Applied Mathematics, University of Western Ontario, London, ON, Canada}
\maketitle
\newpage
\begin{spacing}{1.5}
\begin{abstract}
Clustered Regularly Interspaced Short Palindromic Repeats (CRISPR), linked with CRISPR associated (CAS) genes, play a profound role in the interactions between phage and their bacterial hosts. It is now well understood that CRISPR-CAS systems can confer adaptive immunity against bacteriophage infections. However, the possibility of failure of CRISPR immunity may lead to a productive infection by the phage (cell lysis) or lysogeny. Recently, CRISPR-CAS genes have been implicated in changes to group behaviour, including biofilm formation, of the bacterium \textit{Pseudomonas aeruginosa} when lysogenized. For lysogens with a CRISPR system, another recent experimental study suggests that bacteriophage re-infection of previously lysogenized bacteria may lead to cell death.  Thus CRISPR immunity can have complex effects on phage-host-lysogen interactions, particularly in a biofilm.  In this contribution, we develop and analyse a series of models to elucidate and disentangle these interactions.  From a therapeutic standpoint, CRISPR immunity increases biofilm resistance to phage therapy.  Our models predict that lysogens may be able
to displace CRISPR-immune bacteria in a biofilm, and thus suggest strategies to eliminate phage-resistant biofilms.
\end{abstract}
\end{spacing}
\newpage
\begin{table}[h!]
\begin{spacing}{1.5}
\centering
{\large \caption{ NOMENCLATURE}\label{table:table1}}
\begin{tabular}{ |c|l|c| } 
\hline
Parameters & Description & Dimensions \\
\hline
$H$ & Density of wild-type (non-lysogenic) bacteria & cells cm$^{-2}$ \\ 
$H_L$ & Density of lysogens & cells cm$^{-2}$ \\ 
$C_S$ & Density of bacteria with CRISPR-immunity/spacer& cells cm$^{-2}$ \\ 
$C_L$ & Density of lysogens with CRISPR system & cells cm$^{-2}$ \\ 
$V$ & Density of phage & phage cm$^{-2}$ \\ 
$t$ & Time & hr \\ 
$K$ & Carrying capacity of bacteria in biofilm & cells cm$^{-2}$ \\ 
$\alpha$ & Prophage induction & hr$^{-1}$ \\ 
$r$ & Bacterial growth rate & hr$^{-1}$ \\ 
$b$ & Burst size & phage cell$^{-1}$ \\ 
$\beta$ & Adsorption rate & cm$^{2}$ phage$^{-1}$ hr$^{-1}$ \\ 
$\gamma$ & Phage loss rate & hr$^{-1}$ \\ 
$\theta$ & Adhesion rate per cell & hr$^{-1}$ \\ 
$\eta$ & Slouging off rate of non-lysogens & hr$^{-1}$ \\ 
$\eta_L$ & Sloughing off rate of lysogens & hr$^{-1}$ \\ 
$\phi$ & Planktonic bacteria forming biofilm & cells cm$^{-2}$ hr$^{-1}$ \\ 
$p_L$ & Probability of lysogenization & $-$ \\ 
$p_F$ & Probability of CRISPR failure & $-$ \\ 
$p_D$ & Probability of cell death & $-$ \\ 
\hline
\end{tabular}
\end{spacing}
\end{table}
\newpage
\begin{spacing}{1.5}
\section{Introduction}
The co-existence of bacteria and bacteriophage has been of prolonged interest to evolutionary biologists \cite{abedon_bacteriophage_2008}. Temperate phage are of particular interest as these viruses can reproduce either through the lytic cycle, causing bacterial cell death, or the lysogenic cycle, allowing both phage and host to survive \cite{abedon_phage_2009, barksdale_persisting_1974, echols_developmental_1972, weinbauer_ecology_2004}.  An important factor in this co-existence
is a mechanism known as Clustered Regularly Interspaced Short Palindromic Repeats (CRISPR) that not only provides adaptive immunity against viral infections but also helps to protect the bacterial genome from foreign mobile elements such as  phage and plasmids \cite{bhaya_crispr-cas_2011,sorek_crispr-mediated_2013}.
\par
The building blocks of CRISPR systems were first identified thirty years ago, when inter-spaced DNA repeats were found in specific regions of the \textit{E. coli} genome \cite{ishino_nucleotide_1987}. Research efforts quantified the variation in these sequences, their lengths and positions, in bacterial and archaea genomes \cite{bachellier_bacterial_1997, belkum_short-sequence_1998, lupski_chromosomal_1996} before the name CRISPR was first applied \cite{jansen_identification_2002}. CRISPR systems have now been identified in nearly 45\% of bacterial strains and have been classified into three types that are further divided according to the phylogeny of the CRISPR sequences and their CRISPR-associated (CAS) genes \cite{bhaya_crispr-cas_2011}. Moreover, multiple CRISPR-CAS systems exist in several bacterial strains demonstrating that different types are compatible with each other within a single cell \cite{deng_novel_2013,wiedenheft_rna-guided_2012}.
\par
The CRISPR adaptive defense mechanism proceeds in three main steps \cite{bhaya_crispr-cas_2011}. Firstly, the CRISPR-CAS system requires a specific sequence in phage DNA known as the protospacer that is acquired as a spacer into a designated position in the bacterial genome known as the CRISPR-locus, which lies next to the CAS genes \cite{barrangou_crispr_2007, garneau_crispr/cas_2010}. This acquisition is made possible by means of a short conserved sequence, present in the vicinity of protospacers, known as the protospacer adjecent motif (PAM) \cite{deveau_crispr/cas_2010, mojica_discovery_2013}.  CRISPR-loci have the capacity to store hundreds of spacers in the form of an array, while each spacer in the array is surrounded by short palindromic repeats known as CRISPR-repeats \cite{Kawaji_Exploration_2008}. In the second step, the CRISPR-locus is transcribed and expressed as a single long mRNA that is cleaved into a single spacer and partial repeat known as CRISPR-RNA (crRNA) \cite{hale_rna-guided_2009}. Depending upon the type of CRISPR system \cite{bhaya_crispr-cas_2011,rollins_mechanism_2015}, crRNA is associated with a CAS protein which provides a strong response against any RNA/DNA sequences matching the protospacer and begins to cleave these viral sequences with the help of CAS enzymes; this third step is known as interference. In the case of subsequent infections by the same type of phage, the crRNA associated with the CAS protein promptly responds and cleaves the phage genome. However, the presence of a limited number of crRNA/CAS complexes inside the cell can lead to CRISPR failure in the case of multiple simultaneous viral infections. Moreover, small variations in PAM can also lead to immune failure resulting in phage infection. In addition, rapidly mutating phage can escape this bacterial immunity by varying the protospacer or PAM sequences. A balance between phage diversification and CRISPR immunity can result in a stable co-existing community \cite{andersson_virus_2008, held_crispr_2010, tyson_rapidly_2008}.
\par
Biofilm formation is another protective mechanism used by bacterial colonies.  For example,  \textit{P. aeruginosa} preferentially exhibits an anaerobic biofilm mode of growth at human body temperature \cite{vandervoort_plasma-mediated_2014, yoon_pseudomonas_2002}. In particular, when this pathogen colonizes wound infections (e.g. suppurative or purulent bacterial infections) or the lungs of cystic fibrosis patients \cite{costerton_bacterial_1999,singh_quorum-sensing_2000}, it forms micro-colony structures which exhibit extremely high levels of antibiotic resistance \cite{costerton_bacterial_1999}. Bacterial cells undergo profound phenotypic changes during the transition from free-floating planktonic to biofilm-associated cells \cite{otoole_biofilm_2000}.
\par
 New advancements in phage therapies are considered to be a promising way to eradicate some otherwise untreatable bacterial infections \cite{miller_mathematical_2014, pires_phage_2015}. Although the therapeutic effect of a virulent phage is typically more effective than a temperate phage \cite{skurnik_phage_2006}, virulent phage therapy is not always possible and therefore the use of temperate  phage is sometimes inevitable \cite{capparelli_experimental_2007,paul_lysis-deficient_2011,seed_experimental_2009}. In particular, nearly 100\% lysogenization can be achieved for \textit{P. aeruginosa} when exposed to some
 viruses, \cite{kutter_bacteriophages:_2004}, but these lysogens are normally immune to super-infection and plaque formation by the same phage \cite{webb_cell_2003}.
\par
Recent studies have revealed an effect of the type-I CRISPR-CAS system on the regulation of group behaviours in one strain of \textit{P. aeruginosa}, PA14 \cite{heussler_clustered_2015, zegans_interaction_2009}. In particular, it has been demonstrated that PA14 is unable to form biofilm and loses swarming motility when CRISPR-CAS identifies a specific protospacer and PAM in a prophage sequence in the bacterial genome \cite{heussler_clustered_2015}. 
This result demonstrates that the CRISPR-CAS system, in addition to its role as an adaptive immune response, can regulate the genomic content of bacteria and help protect bacterial colonies from lysogenization; in fact, very few prophage are observed in \textit{P. aeruginosa} genomes that carry CRISPR systems \cite{edgar_escherichia_2010, labrie_bacteriophage_2010}. The regulation of bacterial genomic content by means of the CRISPR-CAS system has been found in other bacterial strains as well, in which not only is lysogeny prevented, but also existing prophage is targeted, typically resulting in cell death due to genomic breakdown \cite{edgar_escherichia_2010}. This self-destruction of lysogens may protect colonies by reducing the chance of prophage induction.
\par
The effects of CRISPR-CAS systems on bacteria-phage interactions have been studied theoretically using a range of mathematical modelling approaches \cite{childs_multiscale_2012,haerter_spatial_2012,heilmann_sustainability_2010, held_crispr-cas_2013, koonin_evolution_2015}. To date, these models have been developed for virulent phage assuming that CRISPR systems function as adaptive immune systems, without affecting other cellular processes such as biofilm formation \cite{heussler_clustered_2015, zegans_interaction_2009} or the cleavage of prophage \cite{edgar_escherichia_2010}. In parallel with these efforts, a number of models have been developed to study the ecological effects of bacterial group behaviours, particularly biofilm formation \cite{ballyk_biofilm_2008,bester_planktonic_2009,emerenini_mathematical_2015,freter_survival_1983,jones_bacteriophage_2011,masic_persistence_2012}. Modelling efforts directed toward phage infection in biofilms are relatively rare \cite{abedon_bacteriophage_2008, stewart_population_1984}. 
\par
The goal of this work is to investigate the effects of a temperate phage infection in a bacterial biofilm.  Our approach is to develop a series of models that help to isolate and disentangle the roles of bacterial hosts, either with or without a CRISPR system, lysogens, and phage. 
Not surprisingly, our models predict that CRISPR-immune bacteria in a biofilm, such as {\it P. aeruginosa} in cystic fibrosis, will be difficult to eradicate by phage therapy.  However our results suggest that lysogens without a functioning CRISPR system may in some parameter regimes be able to invade and dominate the biofilm, offering a possible avenue for therapy.
\section{Model}
\subsection{Model Formulation}
\begin{figure}[h!]
\centering
\caption{Bacterial populations considered in the models. $H$ represents the population of wild-type bacteria, $H_L$ represents the population of bacteria with prophage (lysogens), $C_S$ is the population of bacteria with a CRISPR-CAS system, $C_L$ are lysogens with a CRISPR-CAS system. The bacterial populations $H$ and $H_L$ are considered in Model 1, $C_S$ and $C_L$ in Model 2 and $H_L$ and $C_S$ in Model 3.} \label{fig:flow_diagram}
\includegraphics[trim = 10mm 20mm 10mm 20mm, scale=.5]{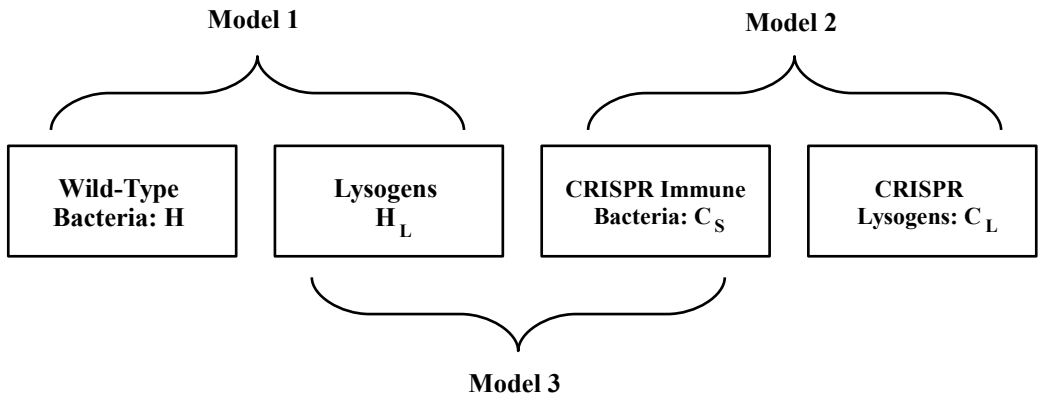} 
\end{figure}

 We develop a series of three models to investigate the effect of infection by temperate phage on a bacterial biofilm, for a host species either with or without a CRISPR-CAS system. Motivated by predominantly single species biofilms as seen in \textit{P. aeruginosa} \cite{heussler_clustered_2015}, all of the models consider cells of a single bacterial species within a biofilm. Each model has two bacterial populations and a population of bacteriophage as described in the Figure \ref{fig:flow_diagram}. In the first model, the population dynamics of host cells, $H$, lysogenized host cells, $H_L$ and phage, $V$, are studied in the absence of a CRISPR system. In the second model, a CRISPR system is introduced; we consider CRISPR-immune host cells, $C_S$, and lysogens with a CRISPR system (but no phage-specific spacer) $C_L$ (Model 2). In the third model, CRISPR-immune bacteria $C_S$ are considered along with lysogens without a CRISPR system. The first case of Model 3 describes the biofilm in isolation whereas in the second case a constant population of planktonic (free) lysogens $\hat{H}_L$ in the environment may join the biofilm. The common parameters of all three models are described as follows.
 \par
 The bacterial populations are modelled as cell densities per unit area of biofilm, cells/cm$^2$. These populations can increase logistically with a maximum growth rate $r$, but are limited by a fixed number of available attachment sites in the biofilm matrix, given by carrying capacity $K$ cells/cm$^2$. The bacteria leave the biofilm with sloughing off rate $\eta$ that may differ for CRISPR and non-CRISPR bacteria. We assume that the virus is able to diffuse fairly freely through biofilm channels, yielding mass action attachment kinetics. Thus, $\beta H V$ gives the number of adsorption events per unit time, and infected bacterial cells typically undergo lysis, producing $b$ daughter phage. The increase in the bacteriophage population is regulated by the rate of phage dissolution, $\gamma$; it is assumed that this phage loss includes phage particles diffusing from the biofilm. We also impose the standard assumption that the loss of phage via productive infection, $\beta H V$ is negligible compared to the overall clearance rate $\gamma V$. Using these parameters, three systems of ordinary differential equations (ODEs) are analysed below to predict which populations will persist at equilibrium for realistic parameter regimes. \\
 \subsection{Model 1}
 The first model is comparatively simple and based on the classical phage-bacteria interaction when there is no CRISPR system in the bacterial cells. Two bacterial populations are considered, the wild-type $H$ and lysogens $H_L$, along with a phage population $V$. The phage can attach to wildtype cells with adsorption rate $\beta$, affecting the host cell in one of two possible ways: (1) host cells $H$ are lysogenized with probability $p_L$, giving rise to lysogens $H_{L}$ or (2) host cells are lysed with probability $(1-p_L)$ which results in the production of a burst of $b$ phage.
 \par
 In lysogens, prophage induction is possible at rate $\alpha$ which again results in cell lysis with burst size $b$. We assume lysogenization confers complete immunity to the virus, that is, $H_L$ cannot be infected. The model is represented by the following set of ODEs,
 \begin{subequations}\label{eq:1}
  \begin{align*}
  \frac{dH(t)}{dt} &= r \left(1-\frac{H+H_L}{K}\right) H-\beta H V - \eta H \\ \\
  \frac{dH_{L}(t)}{dt} &= r \left(1-\frac{H+H_L}{K}\right) H_{L} + p_{L} \beta H V - \alpha H_{L} - \eta H_{L} \\ \\
  \frac{dV(t)}{dt} &= b \left(1-p_{L}\right) \beta H V - \gamma V + b \alpha H_{L}
  \end{align*}
  \end{subequations}
  The purpose of this model is to study criteria for the stable existence of lysogens at equilibrium. This model clearly shows four steady states of which the first is the trivial equilibrium (TE$_1$), in which all populations are zero. This is possible when the sloughing off rate $\eta$ is sufficiently high to eradicate all bacteria from the biofilm. The second equilibrium is the phage free equilibrium (PFE$_1$) in which only wild-type bacteria survive, given by
  \begin{equation} \label{eq:2}
   \begin{bmatrix} H \\ H_L \\ V \end{bmatrix}_{\text{PFE}_1} = \begin{bmatrix} K(r - \eta) \\ 0 \\ 0 \end{bmatrix}
   \end{equation}
    Clearly the existence criterion for steady state PFE$_1$ is $\eta<r$.  The third equilibrium is the lysogenic equilibrium (LE$_1$) in which lysogens survive but the wild-type bacteria die out. These lysogens produce prophage by induction which results in the survival of the phage population. However, these phage have no effect on the bacterial population since lysogens cannot be infected. The equilibrium LE$_1$ is given by
  \begin{equation}\label{eq:3}
   \begin{bmatrix} H \\ H_L \\ V \end{bmatrix}_{\text{LE}_1} = \begin{bmatrix}
     0 \\K \frac{r-(\eta+\alpha)}{r} \\ \frac{R_{_\alpha}}{r}\left(r-(\eta+\alpha)\right) \end{bmatrix}
  \end{equation}
   where $R_{_\alpha}=\sfrac{K b \alpha}{\gamma}$. The existence condition for LE$_1$ is $\eta<r-\alpha$ (note that PFE$_1$ also exists when this condition holds). Finally, there is an all existing equilibrium (AEE$_1$) in which all three populations are non-zero, given by,
   \begin{equation}\label{eq:4}
      \begin{bmatrix} H \\ H_L \\ V \end{bmatrix}_{\text{AEE}_1} = \begin{bmatrix}
       \frac{\bar{V} K p_L}{R_{_\beta}(\alpha - \bar{V})(1-p_L))} \\
       \frac{K}{R_{_\beta} (1-p_L)} \\
       \frac{K\bar{V}}{\beta} \end{bmatrix}
   \end{equation}
   where $\bar{V}=(r-\eta)-\sfrac{r}{R_\beta}$ and $R_{_\beta}=\frac{K b \beta}{\gamma}$. The existence condition for AEE$_1$ is
   \begin{equation*}
   r\left(1-\frac{1}{R_{_\beta}}\right)-\alpha< \eta < r\left(1-\frac{1}{R_{_\beta}}\right).
   \end{equation*}
 The eigenvalues of the Jacobian of each equilibrium provide stability conditions; these are provided in detail in Appendix S1. We find that TE$_1$ is only stable when $r < \eta$, that is, if the sloughing off rate exceeds the maximum growth rate of the biofilm cells. For the local stability of PFE$_1$, the existence condition $\eta<r$ ensures that the first eigenvalue is negative.
 Considering the second eigenvalue, we find the stability condition $\eta > r(1-\sfrac{1}{R_{_\beta}})$ for PFE$_1$, which holds when $R_{_\beta}<1$. $R_{_\beta}$ is analogous to a basic reproductive ratio for the phage, and this result implies that when $R_{_\beta}<1$, the disease-free state PFE$_1$ is stable.
 \par
 The stability conditions for LE$_1$ are closely related to the conditions for PFE$_1$ with an additional dependence on the phage induction rate $\alpha$; the condition for LE$_1$ stability is
 $\eta <  r (1-\sfrac{1}{R_{_\beta}})- \alpha$. If we consider $\alpha$ to be sufficiently small, then stability switches between LE$_1$ and PFE$_1$ at $\eta = r(1-\sfrac{1}{R_{_\beta}})$. Moreover, we see that  $R_{_\beta}>1$  must hold for the stability of LE$_1$. The eigenvalues of the Jacobian evaluated at AEE$_1$ are complicated analytical expressions, however, stability will be explored numerically in the next section. The existence and stability criteria for each steady state in Model 1 are summarized in  Table \ref{table:table2}.
 \end{spacing}
 \begin{table}[ht]
 \centering
 \captionsetup{width=0.8\textwidth}
 \begin{spacing}{1}
 \caption{\small Existence and stability conditions for the steady states in Model 1. The abbreviations used in this table are $R_{_\beta}=\frac{Kb\beta}{\gamma}$, TE$_1$: Trivial equilibrium, PFE$_1$: Phage free equilibrium, LE$_1$: Lysogenic equilibrium, AEE$_1$: all existing equilibrium.}\label{table:table2}
  \end{spacing}
 \begin{spacing}{1.5}
 \begin{tabular}{ |c|c|c| }
 \hline
 Steady State & Existence Criteria & Stability Criteria\\ 
 \hline
 TE$_1$ & $-$ & $\eta>r$ \\
 PFE$_1$ & $\eta<r$ & $\eta > r (1-\frac{1}{R_{_\beta}})$ \\ 
 LE$_1$ & $\eta<r-\alpha$ & $\eta <  r (1-\frac{1}{R_{_\beta}})- \alpha$ \\ 
 AEE$_1$ & $r\left(1-\frac{1}{R_{_\beta}}\right)-\alpha < \eta < r \left( 1 - \frac{1}{R_{_\beta}} \right)$ & evaluated numerically \\
 \hline
 \end{tabular}
 \end{spacing}
 \end{table}
 \begin{spacing}{1.5}
 \par
 We note that, for biologically realistic parameter regimes,
 the phage induction rate $\alpha$ is likely to be very small relative to $r$ and $\eta$.  Thus, AEE$_1$ rarely exists, the existence criteria for PFE$_1$ and LE$_1$ are effectively identical, and their stability criteria are complementary.  The population can stably exist in three states: if $\eta>r$ the host cell population is not sustainable and the trivial equilibrium is stable.  If $\eta<r$, the stability criterion
 for PFE$_1$ determines whether the phage-free or lysogenic equilibrium is stable.  In agreement with intuition, the PFE$_1$ will be stable for $R_{_\beta}<1$.
 
 \subsection{Model 2}
 In this model, we examine how the situation above might differ if the bacterial strains involved have CRISPR immunity.  We thus assume that both the wild-type and lysogenized cells have a CRISPR-CAS system, and denote these populations $C_S$ and $C_L$ respectively. The first population is assumed to have previously acquired the protospacer of phage DNA; these $C_S$ cells are therefore CRISPR-immune. CRISPR-immune bacteria have the ability to resist future infections by the same phage, however there is a small possibility of failure. Therefore, in the case of phage adsorption by $C_S$, the probability of CRISPR failure $p_{_F}$ leads the cell to lysis.
 \par
 For lysogens with a CRISPR system, we note results from a recent experimental study \cite{edgar_escherichia_2010} which demonstrates that bacteriophage infection of previously lysogenized bacteria may lead to cell death, if the CRISPR-CAS proteins acquire the protospacer of the infecting phage. In this case, the same adsorption rate ($\beta$) is considered for the interaction between the bacteriophage $V$ and lysogens $C_{L}$, however this interaction is lethal to the cell with probability $p_{_D}$. Finally, lysogens have a CRISPR-CAS system which regulates their group behaviour \cite{heussler_clustered_2015} so that they lose the expression of polysaccharides essential for biofilm formation. Therefore, the sloughing off rate in $C_L$ increases ($\eta_L > \eta$). This dynamical system can be written as
 \begin{equation}\label{eq:7}
   \begin{aligned}
   \frac{dC_S(t)}{dt} & = r \left(1-\frac{C_S+C_L}{K}\right) C_S-p_{_F}\beta C_S V - \eta C_S \\ \\
   \frac{dC_{L}(t)}{dt} & = r \left(1-\frac{C_S+C_L}{K}\right) C_{L} - p_{D} \beta C_L V - \alpha C_{L} - \eta_L C_{L} \\ \\
   \frac{dV(t)}{dt} & = bp_{F}\beta C_S V - \gamma V + b \alpha C_{L}
   \end{aligned}
 \end{equation}
 We begin by examining the existence conditions for equilibria of this model. Apart from the trivial equilibrium (TE$_2 = (0,0,0)$), there are four more equilibrium states: the phage free equilibrium (PFE$_2$) in which only $C_S$ survives, the lysogenic equilibrium (LE$_2$) in which $C_L$ and $V$ survive, the CRISPR equilibrium (CE$_2$) in which $C_S$ and $V$ survive and the all existing equilibrium (AEE$_2$) in which all three populations exist. The first three non-trivial equilibria can be written as:
  \begin{equation}\label{eq:8}
   \begin{bmatrix} C_S \\ C_L \\ V \end{bmatrix} = \begin{bmatrix} \frac{K(r- \eta)}{r} \\ 0 \\ 0 \end{bmatrix}_{PFE_2},
   \quad
   \begin{bmatrix}  0 \\K \frac{r-(\eta_L+\alpha)}{R_{_\beta} \alpha p_D + r} \\ R_{_\alpha}\frac{ r-(\eta_L+\alpha)} {R_{_\beta}\alpha p_D + r} \end{bmatrix}_{LE_2},
   \quad
   \begin{bmatrix} \frac{K}{p_{_F} R_{_\beta}} \\0 \\ \frac{ (r-\eta)-\frac{r}{ p_{_F} R_{_\beta}}}{\beta p_{_F}} \end{bmatrix}_{CE_2}.\quad
  \end{equation}
  The existence conditions for the phage free and lysogenic equilibria are similar to those obtained from the previous model and are provided in Table \ref{table:table3}. The equilibrium state for AEE$_2$ is given by
  \begin{equation}\label{eq:9}
  \begin{bmatrix} C_S \\ C_L \\ V \end{bmatrix} = \begin{bmatrix} 
  K\frac {\alpha \left(p_{_D} \left(r-\eta \right) -  p_{_F}\left( r-(\alpha +\eta_L) \right) \right) + \frac {r}{R_{_\beta}} \left( \alpha+\eta_L-\eta \right) }{r \left( p_{_F} \left( \eta_L-\eta \right) +\alpha p_{_D} \right)}  \\ K p_{_F} \left( \alpha+\eta_L-\eta \right)\frac {p_{_F} \left(r \left(1-\frac {1}{R_{_\beta} p_{_F}} \right)- (\alpha+ \eta_L) \right) -p_{_D} \left(r \left( 1-\frac {1}{R_{_\beta} p_{_F}} \right) - \eta \right) } {r \left( p_{_F}-p_{_D} \right) \left( p_{_F} \left( \eta_L-\eta \right) + \alpha p_{_D} \right) } 
    \\
  \frac {\alpha+\eta_L-\eta}{\beta \left( p_{_F}-p_{_D} \right) } 
  \end{bmatrix}_{AEE_2}
  \end{equation}
  The existence conditions for the above equilibrium are quite complex as shown in Table \ref{table:table3}, however one necessary condition is $p_{_F}>p_{_D}$. Note that this is the only equilibrium whose existence depends on the relation between the probabilities of CRISPR failure and cell death.
  \par
  All these conditions admit the possibility of co-existence of equilibria and we thus consider stability conditions from the eigenvalues of the Jacobian, as provided in Appendix S2. 
  In brief, for PFE$_2$ we find an analogous condition to that found in previous model for PFE$_1$; in particular, PFE$_2$ is locally stable under the condition $\eta > r \left( 1- \frac{1} {R_{_\beta} p_{_F}} \right)$. 
   For CE$_2$, the second and third eigenvalues are complex conjugates with negative real parts whenever CE$_2$ exists,
   and we find a necessary condition for the negativity of the first eigenvalue. Similarly, for LE$_2$, the second and third eigenvalues are negative whenever the equilibrium exists but we derive a condition for the negativity of the first eigenvalue.
  Finally, the eigenvalues for the Jacobian matrix of AEE$_2$ are sufficiently complicated expressions that we will use numerical methods to assess stability. Table \ref{table:table3} shows a summary of existence and stability conditions for Model 2.
  \end{spacing}
 \begin{table}[ht]
  \centering
  \captionsetup{width=1\textwidth}
  \begin{spacing}{1}
  \caption{\small Existence and stability conditions of the steady states in Model 2. The abbreviations used in this table are $R_{_\beta}=\frac{Kb\beta}{\gamma}$, TE$_2$: Trivial equilibrium, PFE$_2$: Phage free equilibrium, LE$_2$: Lysogenic equilibrium, CE$_2$: CRISPR equilibrium and AEE$_2$: All existing equilibrium.}\label{table:table3}
  \end{spacing}
  \begin{spacing}{1.5}
  \begin{tabular}{|c|c|c|}
  \hline
  Steady State & Existence Criteria & Stability Criteria\\ 
  \hline
  TE$_2$ & $-$ & $\eta>r$ \\
  PFE$_2$ & $\eta < r$ & $\eta > r (1-\frac{1}{R_{_\beta} p_{_F}})$ \\ 
  LE$_2$ & $\eta < \eta_L < r-\alpha$ & $\eta>r-\left(r-(\alpha+\eta_L)\right)\frac{R_{_\beta}\alpha p_{_F}+r}{R_{_\beta}\alpha p_{D}+r}$ \\
  CE$_2$ & $\eta< r\left(1-\frac{1}{R_{_\beta} p_{_F}}\right)$ & $\eta<r\left(1-\frac{1}{R_{_\beta} p_{_F}}\right)\left(1-\frac{p_{_F}}{p_{_D}}\right)+ \frac{p_{_F}}{p_{_D}}(\alpha+\eta_L)$\\
  AEE$_2$ & $\eta < r-\left(r-(\alpha+\eta_L) \right) \frac{R_{_\beta} \alpha p_{_F} + r}{R_{_\beta} \alpha p_{_D} + r}$ & \\ & $\eta>r\left(1-\frac{1}{R_{_\beta} p_{_F}}\right)\left(1-\frac{p_{_F}}{p_{_D}}\right)$ & evaluated numerically \\ & + $\frac{p_{_F}}{p_{_D}} \left(\alpha+\eta_L\right)$,\,\,\,\, $p_{_F}>p_{_D}$ & \\
  \hline
  \end{tabular}
  \end{spacing}
  \end{table}
  \begin{spacing}{1.5}

We find that the conditions for the stable existence of the trivial and phage-free equilibria are  identical to those described
for Model 1, with $R_{_\beta}$ replaced by $R_{_\beta} p_F$.  Thus again if the sloughing off rate exceeds the growth rate, the biofilm cannot sustain itself.  If the biofilm can sustain itself, we find that the phage population cannot be sustained if $R_{_\beta} p_F < 1$.  
If both the bacterial cells and phage are sustainable, however,
three equilibrium states are possible.  Once again, though, we note that $\alpha$ is very small, and thus the lysogenic equilibrium is rarely stable, since $\eta < \eta_L$ by definition.  The all existing equilibrium again has a possibly
narrow parameter range for existence, and thus the model predicts that for biologically relevant parameter values, the
CRISPR equilibrium is most likely to be observed.  This is illustrated further in the numerical work to follow.

  \subsection{Model 3}
  This model consists of a population of CRISPR-immune bacteria $C_S$ along with non-CRISPR lysogens $H_L$ which continually contribute to the phage population $V$ in the biofilm via induction. The populations and parameters are the same as described for the previous models, with the exception of the biofilm formation rate $\phi$. The idea here is to study the impact of a population of non-CRISPR lysogens $H_L$ on the population of CRISPR-immune bacteria $C_S$. This would allow us to investigate the possibility of using lysogens to deliver phage therapy, with the long-term goal of eradicating the CRISPR-immune bacteria. 
  \par
  Since the lysogens $H_L$ do not have a CRISPR-CAS system, it is possible for planktonic lysogens to express polysaccharides and join the outer layer of the biofilm. The model assumes that planktonic bacteria are present in the environment around the biofilm, allowing planktonic bacteria to form biofilm at a specific rate constant of adhesion while the medium carrying planktonic bacteria flows into and out of the system, including the sloughed off bacteria from the biofilm.  We use $\phi$ to denote the maximum rate of biofilm formation by planktonic bacteria; this rate is reduced by the carrying capacity such that the net biofilm formation rate depends on the attachment sites available in the biofilm.  We assume that bacteria are lost from the biofilm independent of the density of planktonic bacteria; this sloughing off rate is assumed to be the same for both $C_S$ and $H_L$.  The model is represented by the following equations,
      \begin{equation}\label{eq:13}
       \begin{aligned}
       \frac{dC_S(t)}{dt} &= r\left(1-\frac{C_S+H_L}{K}\right) C_S-p_{_F}\beta C_S V - \eta C_S \\ \\
       \frac{dH_{L}(t)}{dt} &= \left(1-\frac{C_S+H_L}{K}\right) (rH_{L}+\phi) - \alpha H_{L} - \eta H_{L} \\ \\
       \frac{dV(t)}{dt} &= bp_{F}\beta C_S V - \gamma V + b \alpha H_{L}
       \end{aligned}
       \end{equation}
  The above system is studied in two cases: (1) when there are no planktonic lysogens in the environment ($\phi = 0$) and (2) when there are lysogens in the environment that may join the biofilm ($\phi > 0$).
  \subsubsection*{Case (a): $\boldsymbol{\phi=0}$}
  When the biofilm formation term $\phi=0$, this model becomes relatively simple, yielding four steady states, three of which are defined in the previous model, i.e. the phage-free equilibrium ($PFE_3$), trivial equilibrium ($TE_3$) and CRISPR equilibrium ($CE_3$) with the population of lysogens $H_L$ in place of $C_L$, while the fourth equilibrium, i.e. the lysogenic $LE_3$, is as defined for $LE_1$ with $C_S=0$ instead of $H=0$.
  \par
  The eigenvalues of the Jacobian evaluated at these equilibria are given in Appendix S3, and the existence and stability conditions are summarized in Table \ref{table:table4}. There are similarities in the stability conditions of each equilibrium of this model with those described for the previous models. 
  In particular the stability conditions for the trivial and phage-free equilibria are the same as previously described for Model 2.  In this case since $\alpha$ is small, the CRISPR equilibrium has a narrow range of stability, and biofilms are thus most likely to exist either at the phage-free or lysogenic equilibria.
 
  \par
  \subsubsection{Case (b):  $\boldsymbol{\phi>0}$}
  This situation changes when planktonic lysogens in the environment are able to join the biofilm.  When $\phi>0$, only two equilibria remain. One is the lysogenic equilibrium (LE$_\phi$) and the other is the all existing equilibrium (AEE$_\phi$). Equilibrium LE$_\phi$ is given by:
  \begin{equation}\label{eq:16}
     \begin{bmatrix} C_S \\ H_L \\ V \end{bmatrix} = \begin{bmatrix}
     0 \\
     \frac{1}{2 r} (K(r-\eta-\alpha)-\phi+\sqrt{(K(r-\eta-\alpha)-\phi)^2+4Kr\phi))} \\
     \frac{R_{_\beta}}{2 K r} (K(r-\eta-\alpha)-\phi+\sqrt{(K(r-\eta-\alpha)-\phi)^2+4Kr\phi))} \\
	 \end{bmatrix}_{LE_\phi},
     \quad
  \end{equation}
  It can be observed that the lysogens and phage population are always positive because the discriminant is always positive and greater than the expression outside the square-root.
  \par
  For the stability of LE$_\phi$, a complex condition arises from the first eigenvalue, as given in Appendix S3. However, the following condition is sufficient to ensure the stability of LE$_\phi$: $K\alpha(r-\eta)-\eta \phi < 0$ which implies that $\phi>\frac{K\alpha (r-\eta)}{\eta}$. This condition is further analysed in the numerical section to follow, in which the parameter space is explored to delineate the region of stability.
 \par
 The second equilibrium state AEE$_\phi$ is given by:
     \begin{equation}\label{eq:18}
     \begin{bmatrix} C_S \\ H_L \\ V \end{bmatrix} = 
     \begin{bmatrix}
{\frac {Kb\alpha \left(r - \eta - D_4\beta p_{_F} \right) - D_4 \gamma r}{b r \left(\alpha - D_4 \beta p_{_F}
 \right) }} \\
 \frac {\phi \left(\eta + \beta p_{_F} D_4 \right)}{r
 \left( \alpha - D_4\beta p_{_F} \right)} \\
D_4
	\end{bmatrix}_{AEE_\phi},
    \quad
    \end{equation}
where $C_4=\gamma r-b \beta p_{_F} \left(K(r-\eta)+\phi \right)$ and $D_4=\frac{\left(-C_4+\sqrt{C_4^2 + 4K b^2 \beta^2 \eta p_{_F}^2 \phi}\right)}{\left(2 K b \beta^2 p_{_F}^2\right)}$. Since $D_4$ is always positive, we have two possible existence conditions: 
\begin{equation*}
\begin{aligned}
\text{(1)}& \,\,\, \alpha - D_4\beta p_{_F}>0 \implies \phi < K \alpha\left( 1 -\frac{r}{\alpha+\eta} \left(1- \frac{R_{_\beta}}{p_{_F}} \right)\right) \,\,\,\, \text{and}
\\
\text{(2)}& \,\,\, K b \alpha \left(r - \eta - D_4\beta p_{_F} \right) - D_4 \gamma r > 0 \implies \phi>\frac {K \alpha \left(r-\eta \right) \left( r - R_{_\beta} p_{_F} \left(r-\eta-\alpha \right) \right) }{ \left( R_{_\beta} \alpha p_{_F}+ r \right) \left(R_{_\beta}\alpha p_{_F}+\eta \right)}.
\end{aligned}
\end{equation*}
Once again the eigenvalues of the Jacobian evaluated at AEE$_\phi$ are not compact expressions, and stability will be explored numerically in the next section.
\end{spacing}
 \begin{table}[ht]
 \centering
 \captionsetup{width=1\textwidth}
 \begin{spacing}{1}
 \caption{\small Existence and stability conditions of the steady states in Model 3. The abbreviations used in this table are $R_{_\beta}=\frac{Kb\beta}{\gamma}$, TE$_3$: Trivial equilibrium, PFE$_3$: Phage free equilibrium, LE$_3$ and LE$_\phi$: Lysogenic equilibrium, CE$_3$: CRISPR Equilibrium and AEE$_3$ and AEE$_\phi$: All existing equilibrium.}\label{table:table4}
  \end{spacing}
 \begin{spacing}{1.5}
 \begin{tabular}{ |c|c|c| }
 \hline
 Steady State & Existence Criteria & Stability Criteria\\
 \hline
 TE$_3$ & $-$ & $\eta>r$ \\
 PFE$_3$ & $\eta<r$ & $\eta > r (1-\frac{1}{R_{_\beta} p_{_F}})$ \\
 LE$_3$ & $\eta<r-\alpha$ & $\eta < r (1-\frac{1}{R_{_\beta} p_{_F}})- \alpha$ \\
 CE$_3$ & $\eta<r\left(1-\frac{1}{R_{_\beta} p_{_F}}\right)$ & $r\left(1-\frac{1}{p_{_F} R_{_\beta}}\right)-\alpha < \eta < r\left(1-\frac{1}{p_{_F} R_{_\beta}}\right)$\\
 LE$_\phi$ & $-$ & $\phi>\frac{K\alpha (r-\eta)}{\eta}$ (sufficient condition) \\
 AEE$_\phi$ & $\phi > {\frac {\alpha \gamma K \left(r-\eta \right) \left( \gamma r-Kb
  \beta {\it p_{_F}} \left(r-\eta-\alpha \right) \right) }{ \left( Kb \alpha \beta p_{_F}+\gamma r \right) \left(Kb \alpha \beta p_{_F}+\eta \gamma \right)}}$ & \\ & $\phi < K \alpha\left( 1 -\frac{r}{\alpha+\eta} \left(1- \frac{\gamma} {Kb\beta p_{_F}}\right)\right)$ & evaluated numerically\\
 \hline
 \end{tabular}
 \end{spacing}
 \end{table}
 \begin{spacing}{1.5}
 Model 3 has been developed to investigate the possibility of using lysogenized bacteria as a delivery mechanism in phage therapy.  Without the addition of lysogenized bacteria ($\phi = 0$), the most likely stable equilibrium for the realistic parameter values is PFE$_3$, particularly when $R_{_\beta}<\frac{1}{p_{_F}}$; this corresponds to a biofilm composed of bacteria that are immune to the phage.  However the stability conditions for the two equilibria when $\phi>0$ imply that when the rate of biofilm formation by lysogenized bacteria is sufficiently high, the population of CRISPR-immune bacteria may be eliminated from the biofilm.

\section{Numerical Simulations}
Numerical simulations for the above models have been performed to illustrate the impact of parameter values on the existence and stability of the steady states. In addition, the eigenvalues for the Jacobian at two all existing equilibria (AEE$_1$ and AEE$_\phi$) which were not provided analytically are illustrated numerically.
\subsection{Parameter values}
Baseline parameters were obtained through a review of experimental results in the literature as well as parameter values used in previous mathematical modeling studies (Table 2) \cite{abedon_bacterial_2012, childs_multiscale_2012,freter_survival_1983, jones_freter_2003, kutter_bacteriophages:_2004,levin_population_2013}. The rate of bacterial growth $r = 1$ hr$^{-1}$  is the maximum growth rate \cite{abedon_bacteriophage_2008,childs_multiscale_2012} in the logistic growth term; growth slows at higher densities. The carrying capacity of the bacterial population inside the biofilm is the total attachment sites available per unit area. We take $K=5 \times 10^{6}$ cells/cm$^2$ as chosen by Freter \cite{freter_survival_1983} and followed by \cite{ballyk_biofilm_2008} and \cite{masic_persistence_2012}.
\par
We assume lysogens leave the cell population by means of prophage induction at the rate $\alpha=10^{-5}$ hr$^{-1}$, producing $b=200$ phage copies per cell through lysis. Phage can infect bacteria at the rate $\beta = 10^{-7}$ cm$^{2}$ phage$^{-1}$ hr$^{-1}$ and produce $b=200$ phage copies per cell with probability $1-p_L$, where $p_L=0.05$ is the probability of bacteria gaining prophage. The phage loss rate is assumed to be $\gamma=0.05$ hr$^{-1}$ which includes phage dissolution and sloughing off from the biofilm.
\par
Biofilm formation and  sloughing off are frequently discussed in the literature \cite{ballyk_biofilm_2008, freter_survival_1983, masic_persistence_2012}. In this article, the sloughing off rate of bacteria is the same in all models, except that lysogens in Model 2, $C_L$, leave the biofilm more rapidly due to the CRISPR response. The baseline parameter value for the sloughing off rate for wildtype bacteria is taken to be $\eta = 0.1$ hr$^{-1}$ and for lysogens with a CRISPR system is given by $\eta_L = \sqrt{\eta}$ so that $\eta_L > \eta$ when $0 < \eta < 1$. The parameter value for the biofilm formation rate $\phi$ is found from Freter's model by assuming that the density of bacteria present near the biofilm is constant as $K_2 = 10^8$ \sfrac{cells} {cm$^3$}. The adhesion rate given in Freter's model, i.e. $\tilde{\theta} = 10^{-5}$ m$^3$/hr/gram, is converted to $\theta = \tilde{\theta} K_2$ hr$^{-1}$  $\approx$ 5.56 $\times$ 10$^{-7}$/hr by assuming that 1 gram of bacterial mass can contain $N = 1.8 \times 10^{12}$ cells at maximum. Therefore, the population forming biofilm per hour is given by $\phi = \theta K$ cells/cm$^2$/hr, where $K$ is the carrying capacity of the biofilm.
\par
The CRISPR-CAS effects on the wild-type bacteria and lysogens are parametrised so that they reflect biologically plausible dynamics. Since CRISPR immunity is considered highly efficient with a small probability of failure, we take $p_F = 10^{-4}$ \cite{childs_multiscale_2012}. The probability of failure is a key measure in the models, allowing us to study phage infection in CRISPR-CAS bacteria. Failure is possible since the diversity in CRISPR arrays alters the effectiveness of the CRISPR-CAS system. In particular, CRISPR spacers may become less effective over time and therefore older spacers are preferentially replaced \cite{takeuchi_nature_2012,he_heterogeneous_2010}. The probability of death $p_D$ depends on the adsorption rate of lysogens in the presence of CRISPR-CAS, the presence of repressors for the phage and the diversity of CAS proteins inside CRISPR-immune bacteria. In this study, the same adsorption rate is assumed for lysogens and non-lysogens, while $p_D=0.1$ is used to model CRISPR-CAS initiated cell death.
\end{spacing}
\begin{table}[!ht]
\begin{spacing}{1.5}
\centering
\caption{Baseline values for the parameters used in this study}\label{table:table5}
\begin{tabular}{ |c|l|c| } 
\hline
Parameters & values & Source \\
\hline
$K$ & 5 $\times$ 10$^6$ cells cm$^{-2}$ & \cite{freter_survival_1983,ballyk_biofilm_2008,masic_persistence_2012}\\ 
$\alpha$ & 10$^{-5}$ hr$^{-1}$ & \cite{kutter_bacteriophages:_2004,chibani-chennoufi_phage-host_2004}\\
$r$ & 1 hr$^{-1}$ & \cite{abedon_bacteriophage_2008}\\ 
$b$ & 200 phage cell$^{-1}$ & \cite{abedon_bacteriophage_2008}\\ 
$\beta$ & 10$^{-7}$ cm$^{2}$ phage$^{-1}$ hr$^{-1}$ &  \cite{abedon_bacteriophage_2008}\\ 
$\gamma$ & 0.05 hr$^{-1}$ &  \cite{abedon_bacteriophage_2008}\\ 
$\eta$ & 0.1 hr$^{-1}$ & \cite{freter_survival_1983}\\ 
$\eta_L$ & $\sqrt{\eta}$ hr$^{-1}$ & $-$\\
$\phi$ & $2.78$ cells cm$^{-2}$ hr$^{-1}$ & see text\\ 
$p_F$ & 10$^{-4}$ & \cite{childs_multiscale_2012} \\
$p_L$ & 0.05 & $-$\\ 
$p_D$ & 0.1 & $-$ \\ 
\hline
\end{tabular}
\end{spacing}
\end{table}
\begin{spacing}{1.5}
\subsection{Baseline analysis}
\begin{figure}[!ht]
\centering
\includegraphics[scale=.7]{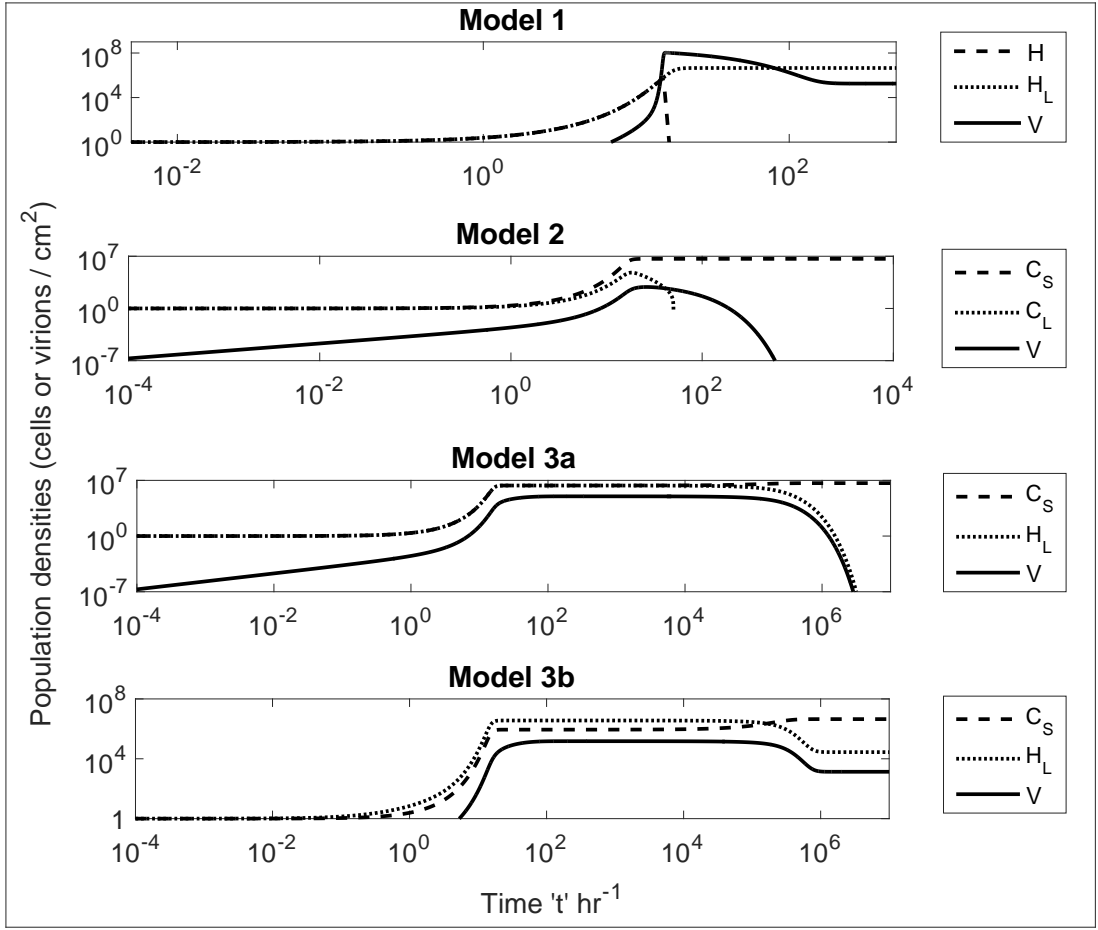}
\caption{Dynamics of phage and bacterial populations inside the biofilm at baseline parametric values. $H$ and $C_S$ represent populations of wild-type bacteria and CRISPR-immune bacteria (dashed lines), $C_L$ and $H_L$ represent lysogens with and without CRISPR systems (dots) and $V$ is the population of bacteriophage (solid lines). Model 1 approaches the lysogenic equilibrium LE$_1$, Model 2 and Model 3 (case (a)) approach the phage-free equilibria (PFE$_2$ and PFE$_3$) whereas Model 3 (case (b)) approaches the all existing equilibrium (AEE$_\phi$).}
\label{fig:Fig2}
\end{figure}

In Figure \ref{fig:Fig2}, the population dynamics of each of these models are illustrated for the same baseline parameter values and initial conditions. The simulations are run over long times to illustrate stable equilibrium states, and in each case the numerical results were validated by comparison with the analytical expressions.  We plot both population densities and time on a log axis, such that initial transients are visible.
\par
In the first model, the baseline scenario produces a stable lysogenic equilibrium (without CRISPR), in which the phage population is maintained in the biofilm through induction. This is a classical model which supports the idea of co-existence of temperate phage and their bacterial hosts.  This result should generalize to the case of many phage types, since bacteria are capable of carrying several prophage as part of their genome.
\par
In the second model, the CRISPR bacteria $C_S$ clearly dominate, since CRISPR immunity makes these cells resistant to the phage. The CRISPR effects on the lysogens $C_L$, i.e. the death of lysogens or increase in the sloughing-off rate, make it possible to eliminate these infected bacteria from the biofilm; their loss is followed by the loss of the phage population. The second model demonstrates that CRISPR systems will disrupt the coexistence observed in Model 1, and predicts that for bacteria with CRISPR systems, stable biofilms are likely to be composed of CRISPR-immune bacteria only.
\par
The populations in the first two models reach stable equilibria
fairly quickly, whereas simulation results for the third model show that the populations take longer to reach their stable states. In the absence of an external source of lysogens (Model 3a), CRISPR-immune bacteria dominate, since the CRISPR system helps the bacteria to maintain the phage-free equilibrium after eliminating non-CRISPR lysogens from the biofilm.  When the external source of lysogens is present (Model 3b), we observe stability of the all existing equilibrium for these baseline parameter values.  This is because a high biofilm formation rate is required to meet the sufficient condition for the stability of LE$_\phi$, thus eliminating the CRISPR cells from the biofilm.  To investigate this further, we turn to parametric analysis to explore scenarios in which lysogens are able to replace CRISPR-immune bacteria in the biofilm. The second goal of the parametric analysis is to find a suitable conditions at which, after the eradication of CRISPR bacteria from the biofilm, the lysogens themselves could be eradicated.
\newpage
\subsection{Parametric analysis}
In this section we illustrate several bifurcations that may occur in biologically realistic parameter regimes.  We first illustrate the changes in equilibria and stability that result from variation in the adsorption rate, $\beta$.  Variation in the prophage induction rate, $\alpha$, phage loss rate, $\gamma$, and sloughing off rate $\eta$ have similar qualitative effects and are shown in Appendix S4. Finally, we examine the interesting case of variation in the biofilm formation rate, $\phi$.  
\subsubsection*{Adsorption rate $\beta$}
\begin{figure}
{\centering
\includegraphics[scale=.7]{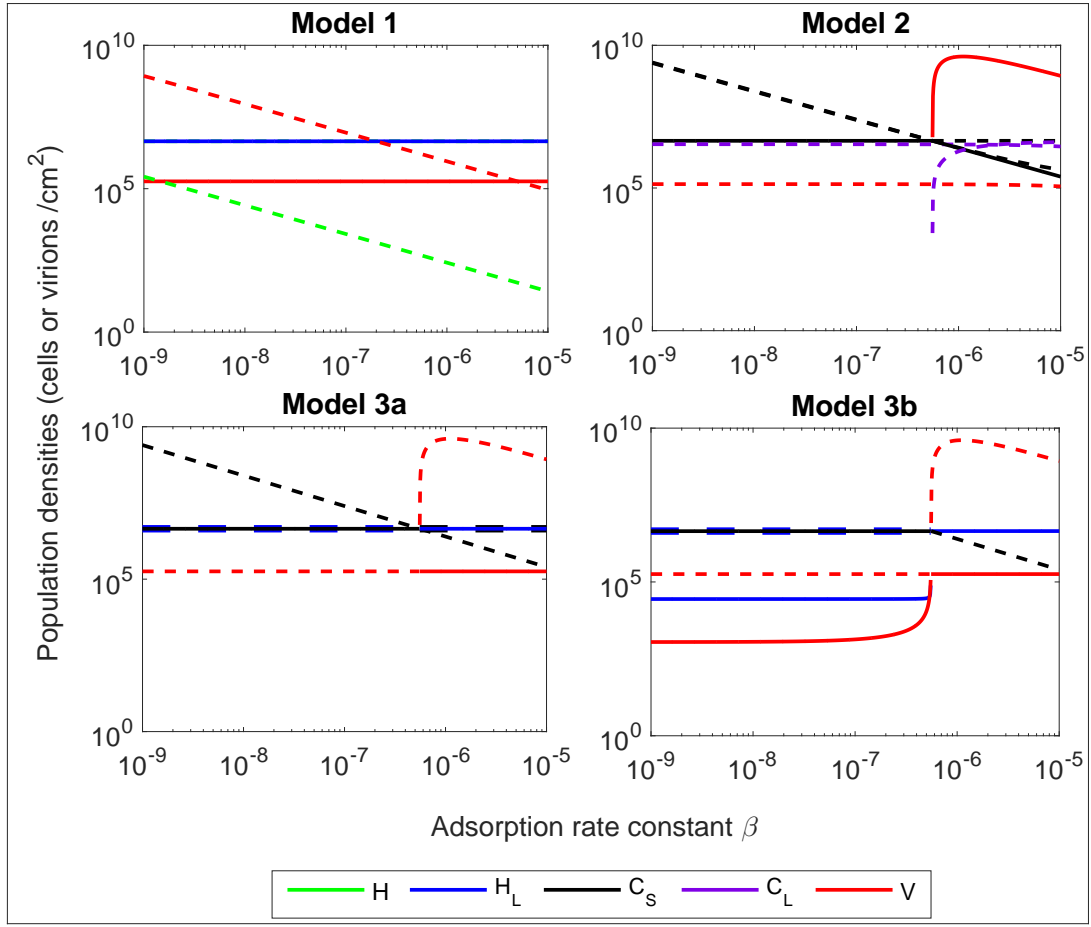}
\caption{Population densities of bacteria and bacteriophage at stable (solid lines) and unstable (dashed lines) equilibrium states against adsorption rate constant $\beta$ in the above models. Each colour represents a unique population while the line width is increased to visualize overlapping populations. Bacterial populations are $H$ (green) and $H_L$ (blue) in Model 1, $C_S$ (black) and $C_L$ (purple) in Model 2 and $C_S$ and $H_L$ in Model 3 whereas the phage population is represented by $V$ (red) in all models.}
\label{fig:beta}}
\end{figure}

Variation in the adsorption rate constant affects the stability of equilibrium states by varying $R_{_\beta}$.  Figure \ref{fig:beta} shows that LE$_1$ remains stable for a wide range of $\beta$ values in Model 1, while the other models each switch stability between two equilibrium states at high rates of adsorption. In Model 2, PFE$_2$ is replaced by CE$_2$ at high adsorption rates.  This is due to the strong immune response of CRISPR bacteria, such that even for high adsorption rates CRISPR-immune bacteria survive, co-existing with the phage.
\par
In Model 3, LE$_2$ is stable at high $\beta$, replacing PFE$_3$ in case (a) and AEE$_\phi$ in case (b).  We note that the biofilm formation rate $\phi$ does not have a strong influence here, since almost same adsorption rate $\beta$ is required in both cases to eradicate CRISPR-immune bacteria. 
\par
Qualitatively, these results demonstrate that for biofilms without CRISPR systems (Model 1), lysogens and phage are predicted to coexist over a wide parameter range.  In contrast, for biofilms with CRISPR (Model 2), the stable equilibrium state is typically dominated by CRISPR-immune bacteria.  If CRISPR-capable lysogens are replaced by lysogens without a functioning CRISPR system (Model 3, a and b), again CRISPR-immune bacteria typically dominate, although for parameter regimes that are extremely favourable to phage reproduction, the lysogenic equilibrium may be stable.  As
shown in Appendix S4, these qualitative conclusions also hold for variations in parameters $\alpha$, $\eta$ and $\gamma$.
\newpage
\subsubsection*{Biofilm formation rate $\phi$}
\begin{figure}[!ht]
{\centering
\includegraphics[scale=.7]{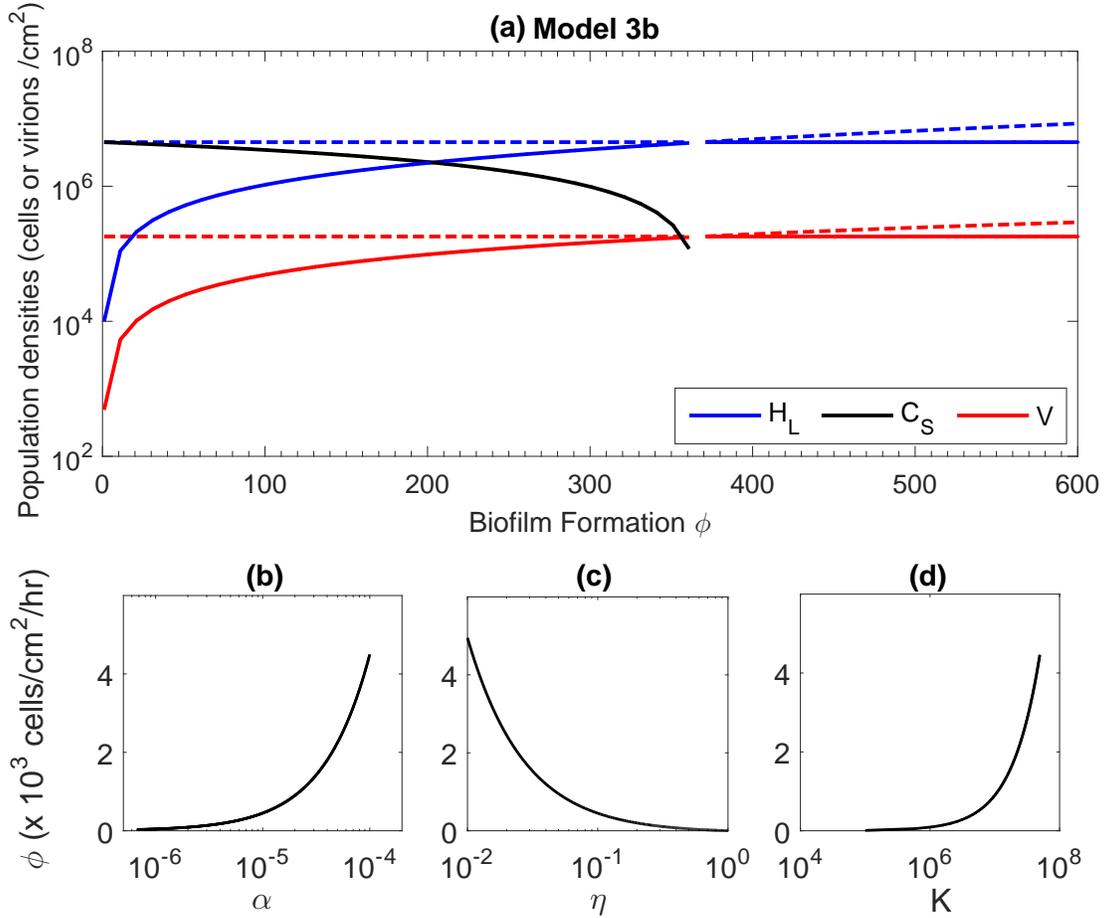}
 \caption{ (a) Population densities of bacteria and bacteriophage at stable (solid lines) and unstable (dashed lines) equilibrium states against phage loss rate constant $\gamma$ in the Model 3 (case (b)). Bacterial populations $C_S$ (black) and $H_L$ (blue) and the phage population $V$ (red) are shown.  Panels (b), (c) and (d) show the critical value of $\phi$ necessary to ensure the stability of the lysogenic equilibrium, versus $\alpha$, $\eta$ and $K$. In each case, the parameter space above each curve corresponds to a stable lysogenic equilibrium (LE$_\phi$) while the space under each curve corresponds to stability of the all existing equilibrium (AEE$_\phi$).}
\label{fig:phi}}
\end{figure}

In Model 3, the rate of biofilm formation by planktonic lysogens is an important parameter that can be regulated either by increasing the density of lysogens near the biofilm or by increasing the adhesion rate.  We illustrate the effect of varying this parameter in panel (a) of Figure \ref{fig:phi}.  As the value $\phi$ increases, the lysogen population increases along with the phage, while the population of CRISPR-immune bacteria $C_S$ decreases in the biofilm, and is eventually eliminated at high values of $\phi$. Though a relatively large formation rate is required to eliminate $C_S$, this threshold value depends on other model parameters.  In particular, the critical
$\phi$ value to ensure the stability of the lysogenic equilibrium in Model 3 (case (b)) is give by $\phi>K \alpha \frac{(r-\eta)}{\eta}$. In the lower three panels of Figure \ref{fig:phi} we plot this threshold against $\alpha$, $\eta$ and the carrying capacity $K$.  Clearly, an increase in the value of $\alpha$ and $K$ require an increase in the value of $\phi$ to make the lysogenic equilibrium stable, whereas an increase in the sloughing off rate $\eta$ reduces the threshold value of $\phi$. 
\newpage
\section{Discussion}
We explore the dynamics of lysogenic phage in a bacterial biofilm, for bacterial hosts both with and without CRISPR immunity.  Classical models of phage-bacteria interactions have previously demonstrated that lysogeny can promote the stable co-existence of bacteria and phage \cite{chibani-chennoufi_phage-host_2004}.  In agreement with these findings, Model 1 explores the baseline conditions under which lysogeny exists in a biofilm, and provides conditions for the stability of a lysogenic equilibrium. The second and third models demonstrate the powerful effect of the CRISPR-CAS system in comparison with the non-CRISPR bacterial population in the first model. The second model predicts that at realistic parameter values, only prophage-free CRISPR-immune bacteria stably survive in a bacterial biofilm.  In rare cases, the CRISPR system stably co-exists with the phage, however the CRISPR system does not allow lysogens to co-exist at this equilibrium.  In the third model, there is no CRISPR system present in the lysogens.  In the absence of an external source
of lysogens (case (a)), CRISPR-immune bacteria are predicted to dominate the biofilm.  However if an external source of planktonic lysogens contributes to the biofilm, CRISPR bacteria may co-exist with lysogens or can even be eliminated from the biofilm.  This last result is of clinical relevance because CRISPR bacteria are highly resistant to phage therapy. Once CRISPR bacteria are removed from the biofilm, phage therapy has a much higher chance of success.
\par
The existence and stability conditions were found analytically to define the parametric regions in which populations exist or remain stable. In order to visualize the population behaviour in the biologically meaningful parameter space, computer simulations were used to verify the stability conditions. In addition, parametric analysis was used to explore realistic parameter regimes,  and to define therapeutic strategies to eradicate CRISPR-immune bacteria from the biofilm.  These results indicate that large magnitude changes in any one baseline parameter value would be required to achieve that objective.  This suggests that means of varying several parameters simultaneously might hold more therapeutic promise.
\par
In the first model, the lysogenic equilibrium is typically stable for realistic parameter values. This equilibrium (LE$_1$) loses stability when $\beta < \frac{\gamma r}{Kb(r-\eta-\alpha)}$. If we consider the sloughing off rate $\eta$ and prophage induction rate $\alpha$ to be negligible compared to the growth rate $r$, then this condition reduces to $\beta < \frac{\gamma}{Kb}$ or $R_{_\beta}<1$ which ensures the stability of PFE$_1$. In between the regions of stability of LE$_1$ and PFE$_1$, there is a small region of length $\alpha$ where AEE$_1$ exists.  Although the biological relevance of this region is arguable, we provide some numerical explorations in Figure \ref{fig:limitcycle}.
\par
The second model predicts that the CRISPR response is sufficiently strong to eliminate lysogens, and therefore CRISPR-immune bacteria dominate the biofilm over a wide range of parameter values. At high infection rates (i.e. $\beta > \frac{\gamma}{Kbp_F}$), assuming the sloughing off rate is negligible compared to the bacterial growth rate, a stable CRISPR equilibrium CE$_2$ exists. This predicts that the phage may exist at high adsorption rates, while lysogen survival is unlikely unless the death probability $p_D$ goes to zero and sloughing off rates $\eta_L$ and $\eta$ become equal. These two conditions would only hold in the unlikely scenario that the CRISPR system has no effect on the lysogens.
\par
The third model explored the case of non-CRISPR lysogens and CRISPR-immune non-lysogens. At baseline parametric values, the lysogens are predicted to go extinct, while the CRISPR-immune bacteria stably exist inside the biofilm.  In some more extreme
parameter regimes, lysogens also persist at equilibrium. The corresponding eigenvalues show that the stability of LE$_3$ and PFE$_3$ have almost complementary conditions, although a small region of bi-stability exists, i.e. $r-\eta<\frac{\gamma r}{Kb \beta p_{_F}}<r-\eta-\alpha$.
\par
Case (b) of Model 3 was specifically designed to explore the possibility of using lysogens to penetrate the biofilm and reduce the population of CRISPR-immune bacteria. A high rate of biofilm formation is necessary to eliminate CRISPR-immune bacteria $C_S$ in this case; both lysogens and phage are always present in the biofilm because of their continuous influx through flow.  Although our model treats only one type of virus and corresponding prophage, an interesting possibility here is that diverse prophage could be introduced via the lysogens joining the biofilm.  In this way,
lysogens could produce a number of different viruses via induction, and this could ultimately reduce the entire biofilm population. Since older CRISPR spacers in the bacteria become less efficient, the possibility of CRISPR failure increases with phage diversity, which helps to eradicate CRISPR-immune bacteria $C_S$.  Once CRISPR bacteria have been eliminated, the biofilm can be treated by classical therapeutic techniques.
\par
\begin{figure}
\centering
\includegraphics[scale=.7]{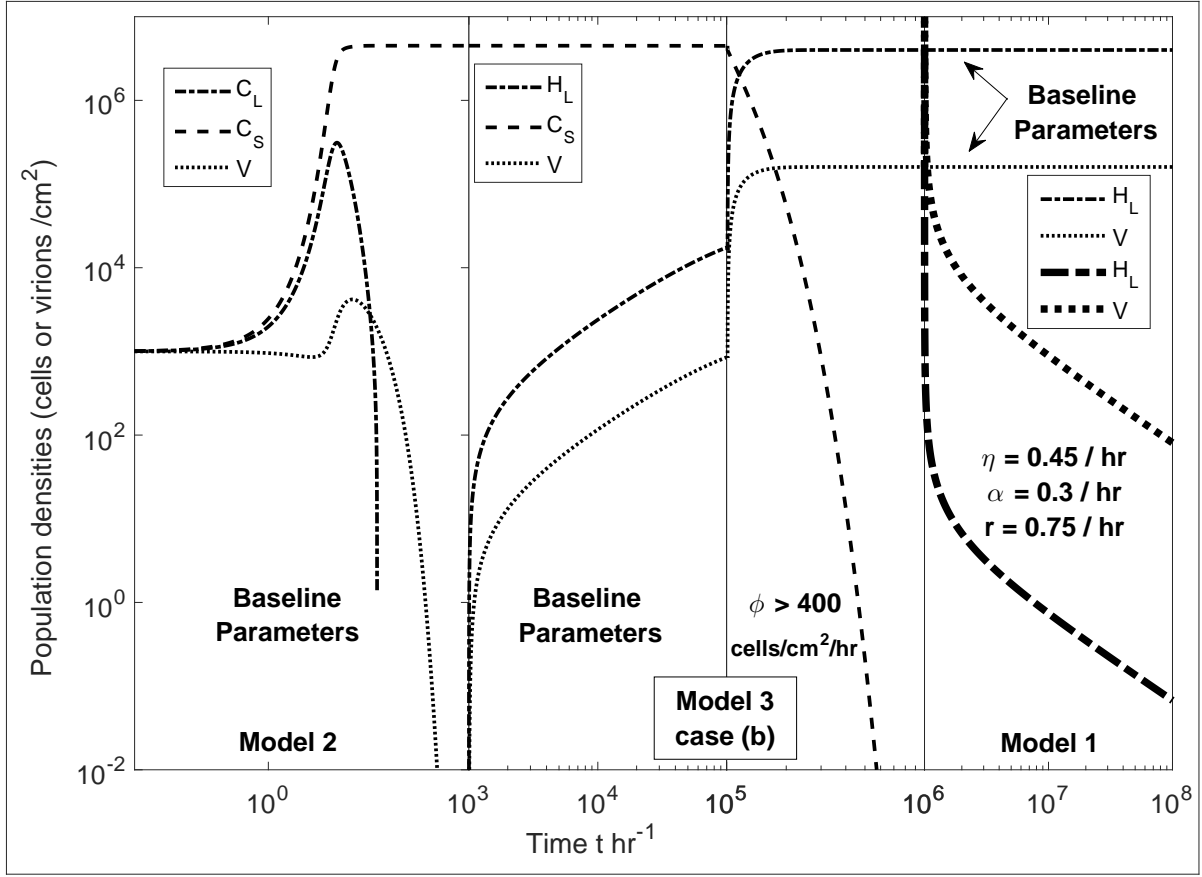}
\caption{Population dynamics of phage-bacteria interaction in Models 2, 3 (case (b)) and 1. Three Vertical lines are drawn to separate the simulations of the models into four panels. Model 2 is simulated for baseline parameters in the first panel, Model 3 (case (b)) is first simulated for baseline parameters and then simulated after an increase in $\phi$ in the second and third panels respectively. In the fourth panel, simulations of Model 1 are presented for baseline parameters (top two curves) and for varied parameters based on possible therapy, i.e. $\eta=0.45$, $\alpha=0.3$ and $r=0.75$, (bottom two curves with increased line-width). The dashed lines represent $C_S$, dash-dot lines represent $C_L$ in Model 2, and $H_L$ in Model 1 and 3, whereas dots symbolize the population density of phage $V$.}
\label{fig:All_model}
\end{figure}

Although the arguments above are highly speculative, the main results of our research are summarized in Figure \ref{fig:All_model} which is divided into four panels (vertical lines). On the left, we use Model 2 to simulate a pathogenic biofilm that is resistant to phage therapy because it consists entirely of CRISPR-immune bacteria.  In the second panel, an external, possibly therapeutic source of planktonic lysogens is applied, and the all existing equilibrium state emerges.  When the concentration of lysogens in the external source is further increased (3rd panel), the CRISPR-immune bacteria are eradicated from the biofilm and the model shows the same behaviour as Model 1 with only lysogens and phage surviving. At realistic baseline parameters, the lysogenic equilibrium emerges.  Once the CRISPR population has been eliminated, the external source of lysogens can be removed and Models 3 and 1 are equivalent.  Variations in a number of parameter values can then be used to eliminate the lysogens (panel 4).  For example, as mentioned above, diversity in the prophages carried by lysogens could produce a number of distinct phage populations inside the biofilm, which could infect bacteria and also increase the prophage induction rate.  Extending the analysis presented here to include multiple phage types is an intriguing possibility for future work.
\newpage
\section*{Acknowledgement}
This work was funded by the Natural Sciences and Engineering Research Council of Canada (NSERC).

\appendix
\section*{Appendices}
\setcounter{figure}{1}
\subsection*{S1\,\,\, Eigenvalues of the Equilibrium States in Model 1}
 The Jacobian of Model 1, evaluated at $PFE_1$ provides the following eigenvalues
 \begin{equation}\label{eq:5}
  \boldsymbol{\lambda}_{PFE_1} = 
  \begin{bmatrix}  -(r-\eta) \\
    -\alpha + \frac{1}{2r}\left(A_1 + \sqrt{ A_1^2 + 4Kb\beta r(r-\eta)\alpha p_L} \right) \\
    -\alpha + \frac{1}{2r}\left(A_1 - \sqrt{A_1^2 + 4Kb\beta r(r-\eta)\alpha p_L}\right)
  \end{bmatrix},
 \end{equation}
 where $A_1=Kb\beta(1-p_L)(r-\eta)-r(\gamma-\alpha)$. Clearly, all eigenvalues are real which implies that there can be no oscillation. Furthermore, the first and the third eigenvalues are always negative for $\eta<r$. Conditions for negativity of the second eigenvalue are discussed in the main text.
 
 The eigenvalues of the Jacobian evaluated at the lysogenic equilibrium $LE_1$ are
     \begin{equation}\label{eq:6}
      \boldsymbol{\lambda}_{LE_1} = 
      \begin{bmatrix} -(r-(\eta+\alpha)) \\ \alpha\left(1-\frac{R_{_\beta}}{r} (r-(\eta+\alpha)) \right) \\
      - \gamma
      \end{bmatrix}
     \end{equation}
 where $R_{_\beta} = \frac{K b \beta}{\gamma}$. When the prophage
 induction rate is high (e.g. $\alpha=0.1$), a narrow range of parameter space allows for the stable existence of AEE$_1$.  Although the biological relevance of this regime is limited, a brief numerical exploration reveals that a stable limit cycle, stable node, or instability are all possible, depending on the initial conditions and the sloughing off rate $\eta$ (see Figure \ref{fig:limitcycle}). 
\begin{suppfigure}
\centering
\includegraphics[scale=.6]{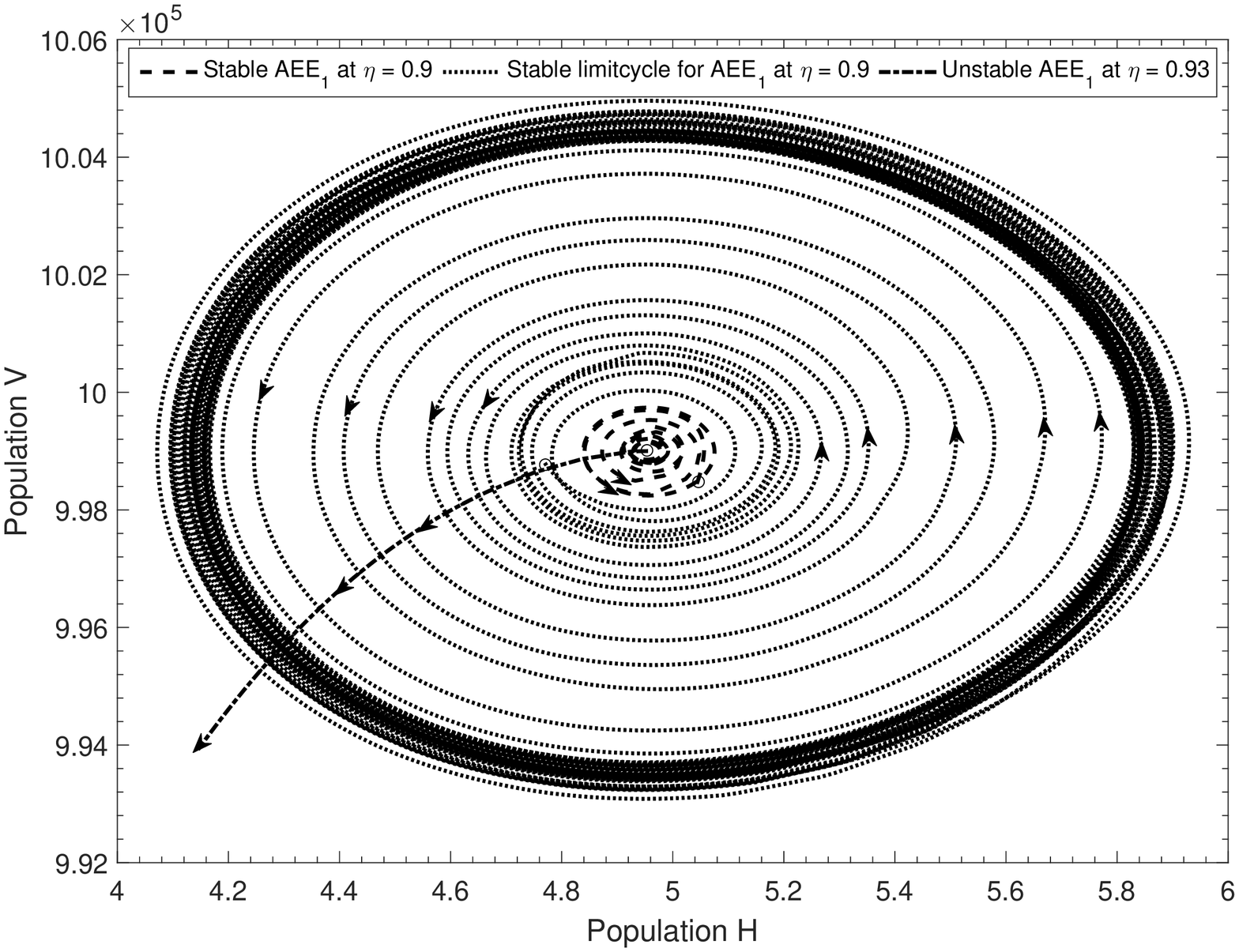}
\caption{Stable and unstable all existing equilibrium at specific parameter values: $\alpha=0.1$, $p_{_L}=0.1$ and $\gamma=0.1$ in Model 1. Bacterial population $H$ and population of virions $V$ are shown in this figure. At $\eta=0.9$, AEE$_1$ shows either a stable equilibrium state (approached by the dashed trajectory in the centre of the figure) or a stable limit cycle (dots, trajectories spiraling outward), whereas at $\eta=0.93$, AEE$_1$ is unstable (dashed-dot line). Each small circle represents the beginning of the curve whereas the arrows represent the direction of the curves.}
\label{fig:limitcycle}
\end{suppfigure}
\subsection*{S2\,\,\, Eigenvalues of the Equilibrium States in Model 2}
The eigenvalues of the Jacobian for Model 2, evaluated at $PFE_2$ are:
    \begin{equation}\label{eq:10}
     \boldsymbol{\lambda}_{PFE_2} = 
     \begin{bmatrix} -(\alpha+\eta_L-\eta) \\
     -(r-\eta) \\
     \frac{R_{_\beta} p_{_F} (r-\eta)-r}{\gamma r}
     \end{bmatrix}
    \end{equation}
  The Jacobian evaluated at CE$_2$ provides the following eigenvalues:
    \begin{equation}\label{eq:11}
	\boldsymbol{\lambda}_{CE_2} = 
	\begin{bmatrix}   
	r\left(1-\frac{1}{R_{_\beta} p_{_F}}\right)\left(1-\frac{p_{_D}}{p_{_F}}\right)- (\alpha+\eta_L) + \eta \left(\frac{p_{_D}}{p_{_F}}\right)
	\\
    -\frac{1}{2}\left(\frac{r}{R_{_\beta} p_{_F}} + \sqrt{ \left(\frac{r}{ R_{_\beta}p_{_F}} \right)^2- 4\gamma \left(r\left(1- \frac{1} {R_{_\beta} p_{_F}} \right) -\eta\right)}\right) \\
    -\frac{1}{2}\left(\frac{r}{R_{_\beta} p_{_F}} - \sqrt{ \left(\frac{r}{ R_{_\beta} p_{_F}} \right)^2- 4\gamma \left(r\left(1- \frac{1} {R_{_\beta} p_{_F}} \right) -\eta\right)}\right) \\
    \end{bmatrix}
    \end{equation}
    Since $\eta<r \left( 1-\frac{1}{p_{_F} R_{_\beta}}\right)$ from the existence condition, the sufficient condition for the negativity of the first eigenvalue is $\frac{p_{_D}}{p_{_F}}\ge 1$, whereas the other two eigenvalues are always negative. Evaluating the Jacobian at the lysogenic equilibrium LE$_2$ provides the following eigenvalues,
       \begin{equation}\label{eq:12}
       \boldsymbol{\lambda}_{LE_2} = 
       \begin{bmatrix}
       \frac{Kb\alpha\beta \left(p_{_D}(r-\eta) - p_{_F} B_2 \right) + \gamma r(\alpha+\eta_L-\eta)}{A_2} \\
       -\frac{\gamma (A_2+rB_2)+\sqrt{\gamma^2 (A_2-rB_2)^2-4 K b \alpha \beta p_D \gamma A_2 B_2}}{2A_2} \\
       -\frac{\gamma (A_2+rB_2)-\sqrt{\gamma^2 (A_2-rB_2)^2-4 K b \alpha \beta p_D \gamma A_2 B_2}}{2A_2}
       \end{bmatrix},
       \end{equation}
     where $A_2=K b \alpha \beta p_{_D} + \gamma r$ and $B_2 = r-(\alpha + \eta_L)$. The last two eigenvalues for LE$_2$ are always negative while the first one requires a condition that is provided in Table \ref{table:table3}.
\subsection*{S3\,\,\, Eigenvalues of the Equilibrium States in Model 3}
 Eigenvalues for the Jacobian of Model 3, evaluated at the CRISPR equilibrium $CE_3$ are given by:
  \begin{equation}\label{eq:14}
    \boldsymbol{\lambda}_{CE_3} = 
    \begin{bmatrix}  
      r(1-\frac{1}{R_{_\beta} p_{_F}} )-(\eta+\alpha) \\
      -\frac{1}{2}\left(\frac{r}{ R_{_\beta}p_{_F}} + \sqrt{\left(\frac{r}{ R_{_\beta}p_{_F}}\right)^2 - 4\gamma \left(\left(r(1- \frac{1} {R_{_\beta} p_{_F}} \right) - \eta\right)}\right) \\
      -\frac{1}{2}\left(\frac{r}{R_{_\beta} p_{_F}} - \sqrt{\left(\frac{r}{R_{_\beta} p_{_F}}\right)^2 - 4\gamma \left(\left(r(1- \frac{1} {R_{_\beta} p_{_F}} \right) - \eta\right)}\right) \\
	\end{bmatrix}
  \end{equation}
  The eigenvalues for the Jacobian of LE$_3$ are given by:
   \begin{equation}\label{eq:15}
      \boldsymbol{\lambda}_{LE_3} = 
      \begin{bmatrix}  
        -\gamma \\
        -(r-(\eta+\alpha)) \\
        -\alpha \frac{R_{_\beta} p_{_F}}{r} \left(r\left(1-\frac{1}{R_{_\beta} p_{_F}}\right) -(\eta+\alpha) \right)
  	\end{bmatrix}
   \end{equation}
    The eigenvalues for the lysogenic equilibrium LE$_\phi$ are found to be
       \begin{equation}\label{eq:17}
       \lambda_{LE_\phi} 
       = \begin{bmatrix}  
   \frac{ R_{_\beta}\alpha p_{_F}}{2 K r} \left(A_4-\sqrt {A_4^2+4K\left(\eta+\alpha\right) }\right) + \frac{1}{2} \frac{B_4 - \sqrt {B_4^2 - 4K\left( K \alpha (r-\eta) - \eta \phi \right)
   }} {K}
   \\
   -{\frac {\sqrt { B_4^2+4K \left( \eta + \alpha \right) \phi}}{K}}
   \\
   -\gamma
     \end{bmatrix},
     \end{equation}
    where $A_4 = -K\left(r- ( \eta+\alpha) \right) +\phi$ and $B_4 =K \left(r +\alpha -\eta \right)+\phi$. The stability conditions for the lysogenic equilibrium LE$_\phi$ in terms of sloughing off rate $\eta$ are:
\begin{equation}
\begin{aligned}
\eta > \frac {1}{K C_4}\left( B_4 +K r C_4 - \frac{1}{2} \frac { B_4  \left(  B_4 - \sqrt{ B_4^{2}-4 R_{_\beta} \alpha p_{_F} \left( r + K\phi C_4^2 \right)} \right) }{R_{_\beta} \alpha p_{_F}} \right) \\
\eta < \frac {1}{K C_4}\left( B_4 +K r C_4 - \frac{1}{2} \frac { B_4  \left(  B_4 + \sqrt{ B_4^{2}-4 R_{_\beta} \alpha p_{_F} \left( r + K\phi C_4^2 \right)} \right) }{R_{_\beta} \alpha p_{_F}} \right)
\end{aligned}
\end{equation}
where $B_4= r+R_{\beta} \alpha p_{_F}$ and $C_4=\frac{ r - R_{\beta} \alpha p_{_F}}{\left( K\alpha+\phi \right)}$.
\subsection*{S4\,\,\, Supplementary Section: Parameter values}
\subsubsection*{Prophage induction rate $\alpha$}
The induction rate was varied to investigate its effectiveness against the CRISPR and non-CRISPR lysogens in the biofilm (see Figure \ref{fig:alpha}). In the case of competition between non-CRISPR bacterial populations, i.e. Model 1, variations in $\alpha$ had no effect; the lysogenic equilibrium remains stable. Likewise, in the presence of a CRISPR system, Model 2, lysogens die out due to the death and sloughing off rates regardless of the value of $\alpha$; this result holds for Model 3a as well.
\par
In case (b) of Model 3, we see that at low rates of prophage induction, i.e. $\alpha < \sfrac{\eta \phi}{K(r-\eta)}$, the lysogenic equilibrium is stable. As the lysogen lifetime is reduced due to increases in prophage induction, the population of CRISPR-immune bacteria also emerges.
\begin{suppfigure}
{\centering
\includegraphics[scale=.7]{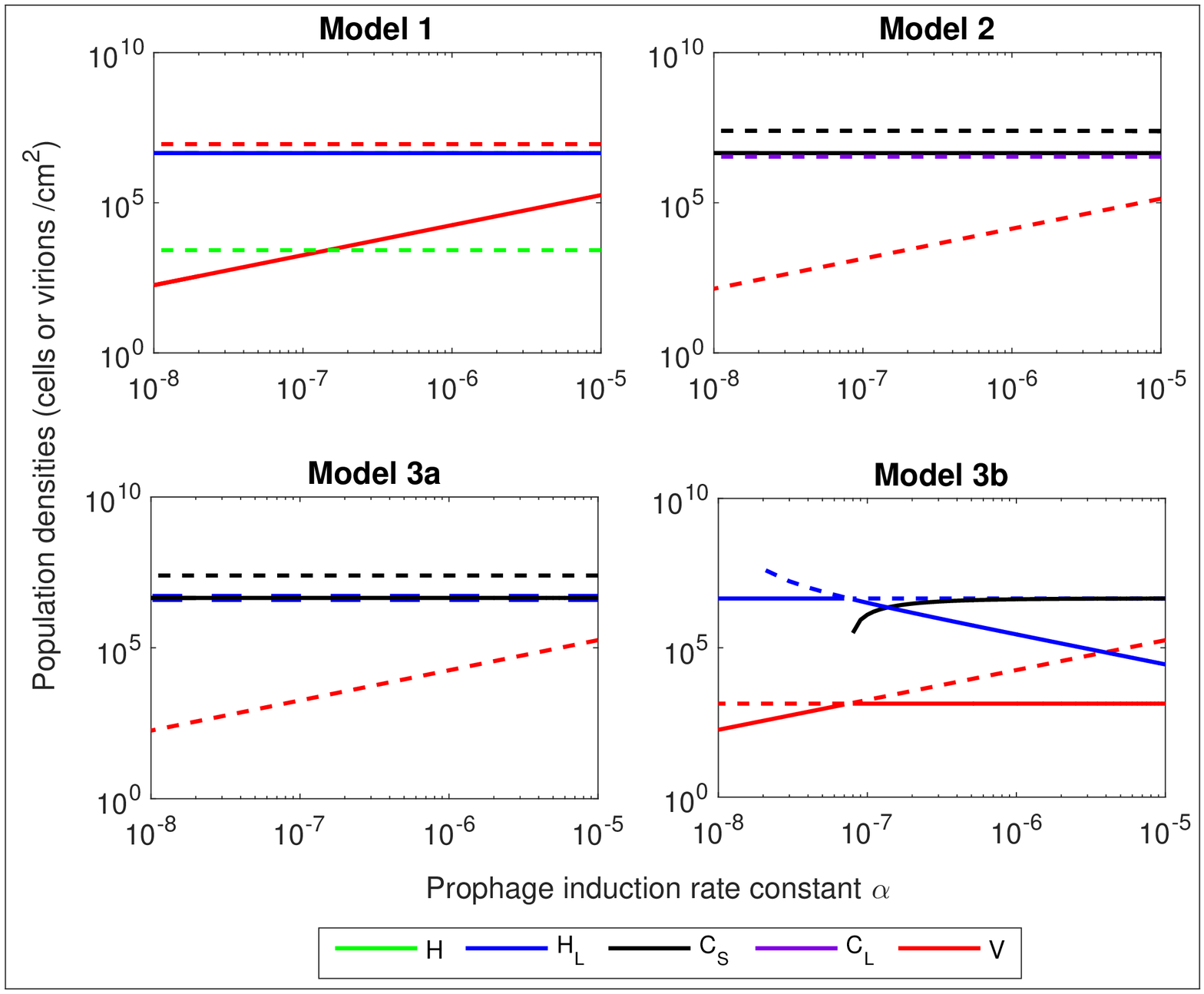}
\caption{Population densities of bacteria and bacteriophage at stable (solid lines) and unstable (dashed lines) equilibrium states against prophage induction rate constant $\alpha$ in the above models. Each colour represents a unique population while the line width is increased to visualize the overlapping behaviours of populations. Bacterial populations are $H$ (green) and $H_L$ (blue) in Model 1, $C_S$ (black) and $C_L$ (purple) in Model 2 and $C_S$ and $H_L$ in Model 3. The phage population is represented by $V$ (red).}
\label{fig:alpha}}
\end{suppfigure}
\newpage
\subsubsection*{Rate of Phage loss $\gamma$}
For the non-CRISPR populations, Model 1, the lysogenic equilibrium is stable at all biologically meaningful values for the rate of phage loss, as shown in Figure \ref{fig:gamma} (top-left plot), and the all existing equilibrium is never stable, even at very high rates of $\gamma$. On the other hand, CRISPR immunity makes it possible for non-lysogens to stably exist at most plausible values of $\gamma$, although for very low $\gamma$ the CRISPR equilibrium (CE$_2$) is stable in Model 2.  In Model 3a, CRISPR bacteria again dominate most of the parameter space (see Figure \ref{fig:gamma}).
\par
In case (b) of Model 3, the external lysogens joining the biofilm do not stabilize LE$_\phi$ significantly; the lysogenic equilibrium is only stable for $\gamma<0.01$. 
For most values of $\gamma$, the lysogens exist along with a population of CRISPR-immune bacteria and a decreasing density of phage in the biofilm, which is an obvious consequence of increasing phage loss.
\begin{suppfigure}
\centering
\includegraphics[scale=.7]{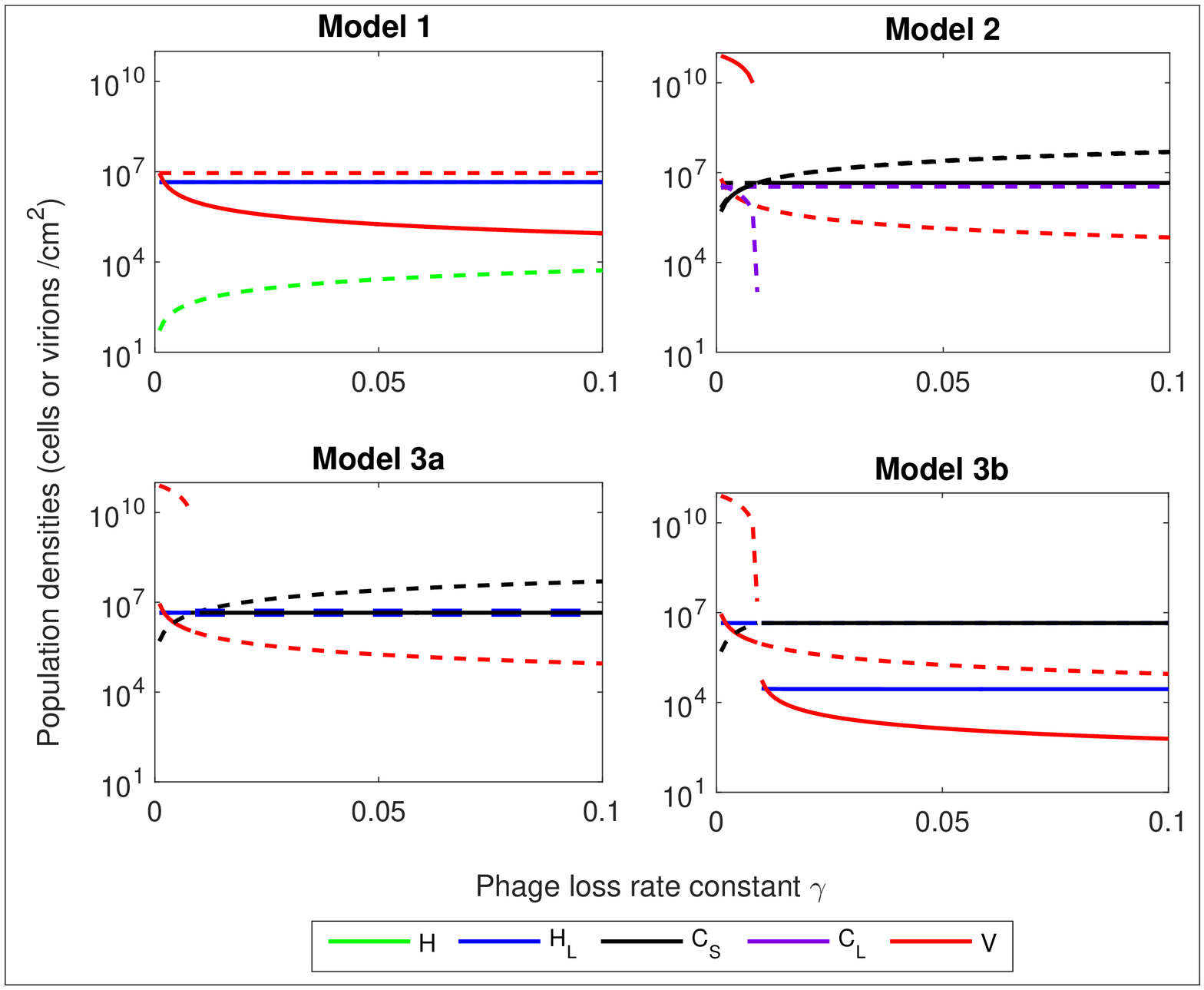}
\caption{Population densities of bacteria and bacteriophage at stable (solid lines) and unstable (dashed lines) equilibrium states against phage loss rate constant $\gamma$ in the above models. Each colour represents a unique population while the line width is increased to visualize the overlapping behaviours of populations. Bacterial populations are $H$ (green) and $H_L$ (blue) in Model 1, $C_S$ (black) and $C_L$ (purple) in Model 2 and $C_S$ and $H_L$ in Model 3. The phage population is represented by $V$ (red).}
\label{fig:gamma}
\end{suppfigure}
\newpage
\subsubsection*{Sloughing off rate $\eta$}
The sloughing off rate is an important parameter in analyzing the stability of each model since it weakens the bacterial population by increasing its rate of loss from the biofilm. The stability of the steady states are very similar to those observed in the above parameter analyses. In general, we see that an increase in the sloughing off rate $\eta$ reduces the population of bacteria and phage in the biofilm, leading at very high rates to the trivial equilibrium (TE). This is because when $\eta$ approaches $r$, all the populations go extinct as shown in Figure \ref{fig:eta}. 
\par
The sloughing off rate $\eta$ affects lysogens either at the same rate as CRISPR-immune bacteria (Model 3a) or at the higher rate (Model 2). Moreover, in case (b) of Model 3, it can be observed that CRISPR-immune bacteria are eliminated from the biofilm before $\eta$ reaches $r=1$, while the lysogens persist because the planktonic lysogens are continuously joining the biofilm. Since there are only two steady states in this case, the parametric value satisfying $\eta>\sfrac{K\alpha r}{K\alpha + \phi}$ is sufficient to eradicate the CRISPR-immune bacteria from the biofilm. Among the parameters involved in the inequality, $\phi$ seems to be an important parameter which can be increased to eradicate $C_S$ even at comparatively low values of $\eta$.
\begin{suppfigure}
\centering
\includegraphics[scale=.7]{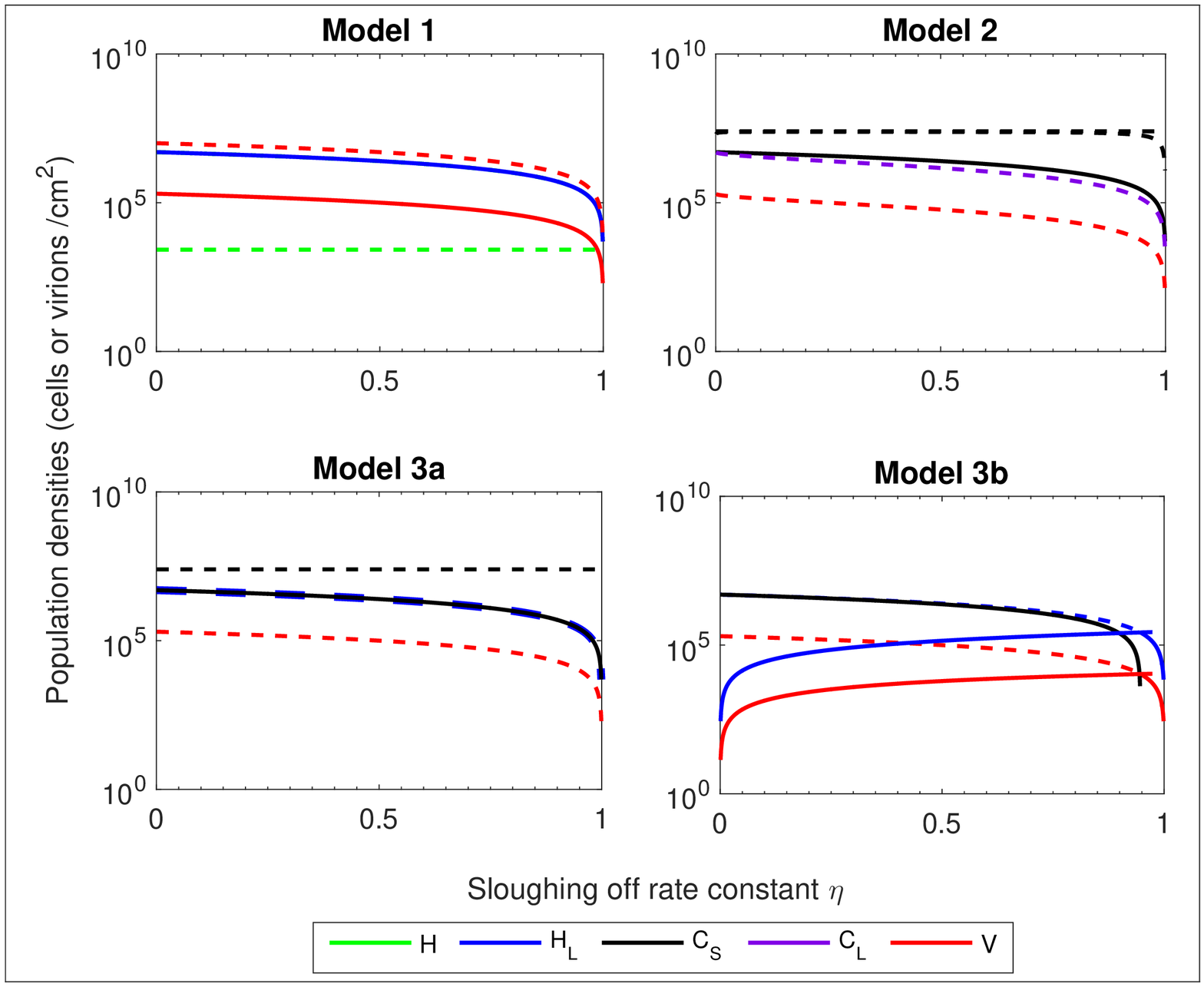}
\caption{Population densities of bacteria and bacteriophage at stable (solid lines) and unstable (dashed lines) equilibrium states against sloughing off rate constant $\eta$ in the above models. Each colour represents a unique population while the line width is increased to visualize their overlapping behaviours. Bacterial populations are $H$ (green) and $H_L$ (blue) in Model 1, $C_S$ (black) and $C_L$ (purple) in Model 2 and $C_S$ and $H_L$ in Model 3. The phage population is represented by $V$ (red).}
\label{fig:eta}
\end{suppfigure}
\newpage
\end{spacing}
\end{document}